\documentclass[preprint,tightenlines,showpacs,
showkeys,nofootinbib,groupedaddress]{revtex4}
\usepackage{epsfig}
\usepackage{amsfonts}
\usepackage{amssymb}
\usepackage{amsmath}
\newcommand{\tmop}[1]{\operatorname{#1}}
\def\issue(#1,#2,#3){{\bf #1}, #2 (#3)} % AIP format

\def\opcit(#1){ {\em op. cit.}, #1}

\def\APP(#1,#2,#3){Acta Phys.\ Polon.\ \issue(#1,#2,#3)}
\def\ARNPS(#1,#2,#3){Ann.\ Rev.\ Nucl.\ Part.\ Sci.\ \issue(#1,#2,#3)}
\def\CPC(#1,#2,#3){Comp.\ Phys.\ Comm.\ \issue(#1,#2,#3)}
\def\CIP(#1,#2,#3){Comput.\ Phys.\ \issue(#1,#2,#3)}
\def\EPJC(#1,#2,#3){Eur.\ Phys.\ J.\ C\ \issue(#1,#2,#3)}
\def\EPJD(#1,#2,#3){Eur.\ Phys.\ J. Direct\ C\ \issue(#1,#2,#3)}
\def\IEEETNS(#1,#2,#3){IEEE Trans.\ Nucl.\ Sci.\ \issue(#1,#2,#3)}
\def\IJMP(#1,#2,#3){Int.\ J.\ Mod.\ Phys. \issue(#1,#2,#3)}
\def\JHEP(#1,#2,#3){J.\ High Energy Physics \issue(#1,#2,#3)}
\def\MPL(#1,#2,#3){Mod.\ Phys.\ Lett.\ \issue(#1,#2,#3)}
\def\NP(#1,#2,#3){Nucl.\ Phys.\ \issue(#1,#2,#3)}
\def\NIM(#1,#2,#3){Nucl.\ Instrum.\ Meth.\ \issue(#1,#2,#3)}
\def\PL(#1,#2,#3){Phys.\ Lett.\ \issue(#1,#2,#3)}
\def\PRD(#1,#2,#3){Phys.\ Rev.\ D \issue(#1,#2,#3)}
\def\PRL(#1,#2,#3){Phys.\ Rev.\ Lett.\ \issue(#1,#2,#3)}
\def\SJNP(#1,#2,#3){Sov.\ J. Nucl.\ Phys.\ \issue(#1,#2,#3)}
\def\ZPC(#1,#2,#3){Zeit.\ Phys.\ C \issue(#1,#2,#3)}

\def\bra {\langle}
\def\ket {\rangle}

\def\del {\partial}
\def\hatlfive {\hat{L_5}}
\def\hatleight {\hat{L_8}}
\def\stwo {\sqrt{2}}
\def\sth {\sqrt{3}}
\def\rat {\varepsilon'/\varepsilon}
\def\ratmath {{\varepsilon'\over\varepsilon}}
\def\beq {\begin{equation}}
\def\eeq{\end{equation}}
\def\bea {\begin{eqnarray}}
\def\eea {\end{eqnarray}}
\def\n {\nonumber}
\def\bc {\begin{center}}
\def\ec {\end{center}}

\begin{document}
\title{Vector  Meson Contributions in $\epsilon'/\epsilon$ }
\author{Anirban Kundu}
\email[Email:]{akundu@cucc.ernet.in}
\affiliation{Department of Physics, University of Calcutta,
92 A.P.C. Road, Kolkata - 700009, India}
\author{Emmanuel A. Paschos}
\email[Email:]{paschos@physik.uni-dortmund.de}
\author{Yu-Feng Zhou}
\email[Email:]{zhou@zylon.uni-dortmund.de}
\affiliation{Universit\"at Dortmund, Institut f\"ur Physik, 
D-44221 Dortmund, Germany}
\date{\today}

\begin{abstract}
  The CP violating parameter $\rat$ is computed using the low-energy dynamics of
  the chiral theory supplemented by vector resonances. The divergent
  contributions coming from strong $\pi$-$\pi$ scattering are tamed by
  vector-meson exchange terms. This amounts to softening the fast growing  high-energy
  behaviour of $\pi$-$\pi$ scattering. The final result for $\epsilon'/\epsilon$
  shows a smooth dependence on the cut-off where low energy dynamics is matched
  with that of QCD.
\end{abstract}
%\preprint{hep-ph/xxxxxx}
%\preprint{CU-PHYSICS-06/2004}
\preprint{DO-TH 03/16}
\preprint{CU-PHYSICS-06/2004}
%\preprint{Prelimilary Version}
\maketitle

%MAIN PART

%\renewcommand{\theequation}{\arabic{section}.\arabic{equation}}
{\bf 1. Introduction.}
The decays $K\to\pi\pi$ are best described by a low energy effective 
Hamiltonian
%eq.(1)
\begin{align}
H=\frac{G_F}{\sqrt{2}}\xi_u
\left\{\sum_{i=1}^8 (z_i(q^2,\mu^2)+\tau y_i(q^2,\mu^2) )Q_i\right\}
\end{align}
with $z_i(q^2,\mu^2)$ and $y_i(q^2,\mu^2)$ being the Wilson coefficients and 
$\xi_q=V^*_{qs}V_{qd}$, $\tau=\xi_t/\xi_u$ . $Q_i$'s are 4-quark operators.  For the definition of the operators
and other notations, see ref.\ \cite{Hambye:PRD} which we closely
follow. Matrix elements for two of these operators, $Q_6$ and $Q_8$,
are most important for the evaluation of $\rat$:
%(2)+(3)
\begin{eqnarray}
Q_6 & = &-2\sum_{q=u,d,s}\bar{s}(1+\gamma_5) q \bar{q}(1-\gamma_5)d,\n\\
Q_8 & = &-3\sum_{q=u,d,s} e_q\bar{s}(1+\gamma_5)q\bar{q}(1-\gamma_5)d,
\end{eqnarray}
where $e_q=(2/3,-1/3,-1/3)$. 
The
QCD corrections included in the Wilson coefficients represent the
short distance terms computed in perturbative QCD.  They depend on $[\ln
(Q^2/\mu^2)]^{\gamma/\beta}$ and to next-to-leading order (NLO) corrections in a
more complicated way.  The numerical values have been tabulated by various
groups \cite{ref1,ref2}. Comparisons of the results show that the various groups
agree with each other but values for the coefficients depend on the
renormalization scheme.  The $\mu$-dependence in the coefficients is expected to
be cancelled by the scale dependence of the matrix elements of the operators
introduced through the upper cut-off in the integrals, and the running strange quark mass.

The matrix elements of the form $\langle \pi\pi|Q_6|K\rangle$ and
$\langle \pi\pi|Q_8|K\rangle$ include tree level contributions and loop corrections.
 These are low energy processes which must be dealt with by
 methods other than QCD. Our method is to use the low energy
chiral theory for calculating tree and loop diagrams and then
match the results with the short distance contribution, i.e.  
the QCD scale $\mu$
is matched with the upper cut-off $\Lambda_c$ appearing in the
chiral loops.  An important criterion for the success of the
calculation is smooth (and weak) dependence of the results on $\mu=\Lambda_c$.

%The method developed in Dortmund-Fermilab \cite{Hambye:PRD,Hambye:NPB} separates 
%the chiral contributions  into
%%\begin{itemize}
%%\item
%(i) factorizable and (ii) nonfactorizable diagrams.
%%\end{itemize}
%Quadratic and other divergences of the factorizable diagrams to
%$O(p^0/N_c)$ were shown to be absorbed into renormalization
%constants. This is so because factorization separates the diquark
%operators (one meson density) whose renormalization confines to
%itself.  Consequently whatever renormalizes a diquark operator
%once in a diagram, it does so in other factorizable diagrams,
%especially to $O(p^0/N_c)$ when the diquark vertices bring no
%momentum dependence. It was
%shown that to $O(p^0/N_c)$ the infinities for $Q_6$ and $Q_8$ are 
%absorbed
%in $\hat{L}_5, \hat{L}_8$. 
In the large $N_c$ approach factorizable and non-factorizable amplitudes are
treated separately \cite{BBG} with the factorizable amplitudes defining
the renormalized coupling constants. In a Dortmund-Fermilab collaboration
\cite{Hambye:PRD}, it was shown that to $\mathcal{O}(p^0/N_c)$ the divergences
in the matrix elements of the $Q_6$ and $Q_8$ operators are logarithmic and
occur in nonfactorizable diagrams.

The numerical results of this approach at $\mathcal{O}(p^0/N_c)$ were
presented in table I of ref.\cite{Hambye:PRD}, which we also adopt
in the present article. 
%
%A result of this procedure is that all remaining divergences,
%which to this order are logarithmic, occur in the nonfactorizable diagrams.
%
The results of the diagrammatic method were reproduced in the 
background-field method \cite{Hambye:NPB}.  Let us denote the nonet
of pseudoscalar meson by the matrix $\Pi=P_a\lambda^a$, 
where $\lambda_a$'s are the usual Gell-Mann matrices; then
it was shown that to $O(p^0/N_c)$
\begin{align}
{\pi_0\over f} = {\pi_r\over F_\pi}\ \ {\rm and}
 \ \ 
{K_0\over f} = {K_r\over F_K}
\end{align}
where $(\pi_0,K_0)$ and $f$ are the bare pion and kaon fields and decay
constants, while $(\pi_r,K_r)$ and $F_{\pi}, F_K$ are renormalized
fields and decay constants, respectively.

%The matrix element for $Q_8$ includes contributions from tree
%diagrams to order $p^0$ and $p^2$ and a loop term of order
%$p^0/N_c$.  In a twofold expansion in terms of momenta and
%loops they comprise all terms to first order.  For the operator
%$Q_6$ the first term of $O(p^0)$ vanishes, so that the expansion
%starts at $O(p^2)$.  Previous calculations \cite{ref1} included
%the $O(p^0/N_c)$ loop and a term to $O(p^2/N_c)$.  The latter
%term introduces a quadratic dependence on the cut-off, which
%cannot be matched with the QCD scale $\mu$.  The large correction
%originates from the process

A large correction in the earlier calculation \cite{Hambye:NPB} originates from
rescattering of the pions, i.e., $K\to \pi\pi\to\pi\pi$ \footnote{The initial
  state interactions are expected to give smaller contributions, which we will
  present in a future publication \cite{longpaper}}
%\vskip -0.5 cm
\begin{figure}[htb]
\begin{center}
\epsfig{file=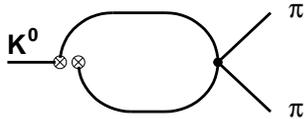,width=7cm}
\caption{Feynman diagram  for $K\to \pi \pi$ with strong final state  interactions.
}\label{contact}
\end{center}
\end{figure}
\noindent 
where the first step involves the weak operators $Q_6$ or $Q_8$ to
$\mathcal{O}(p^2/N_c)$ and the second process is the strong pion-pion
scattering as shown in Fig. \ref{contact}.  The large dependence of the cut-off resides on the contact
$\pi-\pi$ scattering which is known to have a bad high-energy behaviour
violating unitarity and needs to be moderated by some other amplitudes which
restore unitarity.

\begin{figure}[htb]
\begin{center}
\epsfig{file=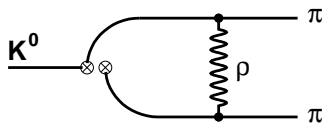,width=7cm}
\caption{Feynman diagram for $K\to \pi \pi$ with a vector-meson exchange.}\label{exchange}
\end{center}
\end{figure}
A standard prescription to restore unitarity is to introduce vector-meson
exchange diagrams.  For the $\pi\pi\to\pi\pi$ scattering we shall use the
contact and the $\rho$ exchange diagrams. We accomplish this by using a chiral
Lagrangian for pseudo scalars and enlarged by the introduction of vector mesons
\cite{Bando,ref5,ref5b}. We extend the calculation of the one-loop diagrams with a strong
vertex with the addition of a $\rho$-exchange diagram.  The $\rho$ is included
to represent the effects of even heavier vector mesons (like $K^*$).  In
addition the pions are in $I=0$ or $I=2$ states and the exchange of
$\rho$-mesons appears only in the $t$-channel, see Fig.\ref{exchange}.
%\vskip -1cm

In order to restore unitarity we shall demand that quadratic
divergences  cancel between the contact and the $\rho$-exchange
diagrams.  It is indeed heartening to note that they come with 
opposite signs, and cancel exactly if the following relation is
satisfied
%eq.(4)
\begin{align}\label{cancellation}
\frac{h^2}{m^2_{\rho}}=\frac{1}{3f^2}\, .
\end{align}
Here $h$ is the $\rho\pi\pi$ coupling strength and $f$ is the 
pion decay constant ($\approx 92 MeV$).  The logarithmic divergences
still remain and should be matched to the QCD logarithms. 
This is our proposal for moderating the high energy growth of $\pi-\pi$
scattering. 

Thus, we calculate the one-loop $K\to\pi\pi$ amplitudes with both contact and
$\rho$-exchange diagrams, demanding that the quadratic divergences cancel betwen
these two sets. The value of $h\simeq 4.8$ obtained from Eq.(\ref{cancellation})
is slightly smaller than the one obtained
from the $\rho$ decay width, but remember that $\rho$ is only a symbolic
representation of all possible vector resonances.  
%This is nothing but
%regularizing the bad $\pi\pi\to\pi\pi$ scattering, contracting one pair of
%$\pi$'s with the $\pi$'s coming out of the weak vertex. 
Since only logarithmic
divergences will be present in the final result, the variation of $\rat$ with
the cut-off $\Lambda$ is expected to be weak. As the weak vertex (with $Q_6$ or
$Q_8$) is common to both the contact and the $\rho$-exchange diagrams, the
cancellation of quadratic divergences is respected by both operators.

%The paper is arranged as follows. In Section 2 we tabulate the necessary 
%formulae and the numerical inputs, and in Section 3 show our results for
%the amplitudes. Section 4 contains the numerical analysis, and in
%Section 5 we summarize and conclude. Some calculational details have been
%relegated to the Appendix \ref{appendixA}. 

%\section{The Framework and input parameters}
{\bf 2. Framework.}
The effective Lagrangian for pseudoscalar mesons relevant for
$K\to\pi\pi$ decay up to ${\cal O}(p^4)$ 
is given by \cite{gasser-leutwyler}:
\bea
{\cal L}_{eff} &=& {f^2\over 4}\left(\bra\del_\mu U^\dag \del^\mu U\ket +
{\alpha\over 4N_c}\bra \ln U^\dag-\ln U\ket^2 + r\bra {\cal M}U^\dag
+UM\ket\right)+r^2H_2\bra M^2\ket\nonumber\\
&{}&+ rL_5\bra \del_\mu U^\dag
\del^\mu U\left( M U + U^\dag M\right)\ket
+r^2 L_8\bra  MUMU + MU^\dag MU^\dag
\ket,
    \label{l-def}
\eea
with $\bra A\ket$ denoting the trace of $A$ and $M={\rm diag}(m_u,
m_d,m_s)$,
$f$ and $r$ are free parameters related to the pion decay constant
$F_\pi$ and to the quark condensate respectively, with $r=-2\bra \bar{q} q
\ket/f^2$. 

The matrix $U$ is given by
\begin{align}
U = \exp(i\Pi/f),
\end{align}
where the pseudoscalar meson nonet $\Pi$ is given by
\begin{align}
\Pi = \lambda^aP_a = \begin{pmatrix}
\pi^0 + {1\over\sth}a\eta + {\stwo\over\sth} b\eta' &
\stwo \pi^+ & \stwo K^+\cr
\stwo \pi^- & -\pi^0 + {1\over\sth}a\eta + {\stwo\over\sth} b\eta' &
\stwo K^0 \cr
\stwo K^- & \stwo \bar{K^0} & -{2\over 3}b\eta + {\stwo\over\sth}a\eta'
\end{pmatrix}
\end{align}
where $\lambda$'s are the usual Gell-Mann matrices, $P_a$ are the pseudoscalar fields,
and
\begin{align}
a = \cos\theta - \stwo\sin\theta,\ \ 
b = {1\over \stwo}\sin\theta + \cos\theta,
\end{align}
$\theta$ being the $\eta-\eta'$ mixing angle. We include $\eta$ and $\eta'$
 contributions in our calculation.
%The operators $Q_6$ and $Q_8$ are given  in terms of the meson fields
%\cite{Hambye:PRD
%\bea
%Q_6 &=& -2f^2r^2\sum_q \huge[ {1\over 4}f^2 U^\dag_{dq}U_{qs} 
%+ U^\dag_{dq} \left(L_5 U\del_\mu U^\dag\del^\mu U + 2rL_8U M
%U + rH_2 M\right)_{qs}\nonumber\\
%&{}& + \left(L_5U^\dag\del_\mu U\del^\mu U^\dag +
%2rL_8U^\dag  MU^\dag + rH_2 M\right)_{dq}U_{qs}\huge]
%+ {\cal O}(p^4),\nonumber\\
%Q_8 &=& -3f^2r^2\sum_q e_q \huge[ {1\over 4}f^2 U^\dag_{dq}U_{qs} 
%+ U^\dag_{dq} \left(L_5 U\del_\mu U^\dag\del^\mu U + 2rL_8UM
%U + rH_2 M\right)_{qs}\nonumber\\
%&{}& + \left(L_5U^\dag\del_\mu U\del^\mu U^\dag +
%2rL_8U^\dag  MU^\dag + rH_2 M\right)_{dq}U_{qs}\huge]
%+ {\cal O}(p^4).
%\eea
It is easy to see that though the operator $Q_6$ vanishes at tree-level
due to the unitarity of $U$, it still has nonzero contributions
at the $\mathcal{O}(p^0/N_C)$ level. The loop expansion of the matrix elements is a series
in $1/f^2\sim 1/N_c$, which follows from the short-distance expansion
in terms of $\alpha_s/\pi \sim 1/N_c$. 

%The factorizable diagrams normalize $f$ to $F_\pi$ and $F_K$, and 
%$L_5$ and $L_8$ to $\hatlfive$ and $\hatleight$. The expressions are to be
%found in \cite{Hambye:PRD}.

%The ratio $\rat$ is given by
%\begin{align}
%\ratmath = {\omega\over\stwo |\varepsilon|}\left( {{\rm Im}A_2\over
%{\rm Re} A_2} - {{\rm Im}A_0\over {\rm Re}A_0}\right),
%\end{align}
%where $A_I$ indicates the $K\to\pi\pi$ amplitude in the isospin $I$ 
%channel, and $\omega = {\rm Re} A_2/{\rm Re} A_0 = 1/22.2$. 

There have been numerous calculations of $\rat$ which try to improve various steps
\cite{epsilonP}.
The expression of $\epsilon'/\epsilon$
can be written in a compact notation as
\begin{align}
\ratmath = {G_F\over 2}{\omega\over |\varepsilon|{\rm Re}A_0} {\rm Im}
\xi_t \left[ \Pi_0 - {1\over\omega}\Pi_2\right], \qquad (\omega=1/22)
\end{align}
with
\begin{eqnarray}
\Pi_0 &=& |\sum_i y_i(\mu) \bra Q_i\ket_0 | \left(1-\Omega_{\eta+\eta'}\right),
\nonumber\\
\Pi_2 &=& |\sum_i y_i(\mu) \bra Q_i\ket_2 |.
\end{eqnarray}
The isospin breaking effect ($m_u\not= m_d$) is taken into account by 
$\Omega_{\eta+\eta'}$. 

%***********
%We include the vector mesons as massive gauge bosons in a theory with  {\em local}
%$SU(3)_L \otimes SU(3)_R$ chiral symmetry \cite{ref5}. The covariant derivative
%is defined through
%\begin{align}
%D_{\mu} U= \partial_\mu U-ig A^L_\mu U+ig U A^R_\mu ,
%\end{align}
%The vector nonet $V$ is defined in a gauge theory with an octect of vector mesons 
%\cite{ref5}
%In the
%present discussion we neglect $\omega$ and $\phi$ mesons; so $\pi$'s are
%replaced by $\rho$'s and $K$'s by $K^*$'s, while analogues of $\eta$ and
%$\eta'$ are absent. 
%The covariant derivative is defined as
%\begin{align}
%%D_\mu P = {1\over\stwo} (\del_\mu - {ih\over 2}\l_aV^a_{\mu})\l_b P^b
%\end{align}
%where the factor $1/\stwo$ is for correct normalization of the kinetic term.
%The three-point vertices are obtained from the term
%where $g$ is the coupling strength, $A^L_\mu=\lambda^a A^L_{a\mu}$ and 
%$A^R_\mu=\lambda^a A^R_{a \mu}$ are the left-handed
%and right-haned gauge fields. The vector and axial-vector fields are defined
%as $V_\mu=(A^L_\mu+A^R_\mu)/2$ and $A_\mu=(A^L_\mu-A^R_\mu)/2$ respectively. 
%With the help of a special gauge transformation, one can eliminate the axial-vector current,
%i.e. in the primed gauge, $A'_\mu=0$ with the remaining vector field becoming the dynamical 
%$\rho_\mu$ field. 
%The local symmetry is explicitly  broken by adding symmetry breaking terms
%such as the mass term of the vector mesons.
% 
%
%
%\begin{align}
%{ih\over 4}(\l_c\l_d\l_b - \l_b\l_d\l_c) P^c V^d_{\mu} \del_\mu P^b
%\end{align}

Our aim is to introduce vector mesons in terms of a Lagrangian which satisfies the low
energy current algebra. One consistent method is in terms of a non-linear chiral Lagrangian with
a hidden local symmetry \cite{Bando}. In this theory the vector mesons emerge as dynamical vector mesons. 
The three point vector-pseudo scalar interaction is given by
\begin{align}%\label{}
\frac{ih}{4}\langle V_\mu(P\partial^\mu P-\partial^\mu P P)\rangle,
\end{align}% 
where $h$ stands for  the vector-pseudoscalar coupling. Some typical vertices of 
$\rho$'s to pseudoscalar mesons are
\begin{align}
\pi^+(p_1)\pi^-(p_2) \rho^0  &: h (p_1-p_2)_{\mu}\epsilon^\mu \nonumber\\
\pi^+(p_1)\pi^0(p_2)\rho^-  &: h (p_1-p_2)_{\mu}\epsilon^\mu \nonumber\\
K^+(p_1)\bar{K}^0(p_2)\rho^-  &: \frac{h}{\sqrt{2}} (p_1-p_2)_{\mu}\epsilon^\mu, \quad etc,
\end{align} 
 which is directly related to the 
$\rho$ decay width:
$\Gamma(\rho)= h^2(|\mathbf{p}_\pi|)^3/(6 \pi m_\rho^2)$, where $\mathbf{p}_\pi$ is the
momentum of final state pions in the $\rho$ rest frame.  With $\Gamma(\rho)=149.2$ MeV,
we find $h=5.95$. We note in passing  that  the Kawarabayashi-Suzuki-Riazuddin-Fayyazuddin 
relation gives 
the value $h=m_\rho/(\sqrt{2} f_\pi)$\cite{KSRF}. Thus the value of $h$ in Eq.(\ref{cancellation})
and the two values in this paragraph differ by small amounts $(\sim 19 \%)$. 
%\end{align}%
%
%
The strong four-point vertices involving pions are obtained from the
first two terms of Eq. ({\ref{l-def}}). The weak vertices are obtained from
the definitions of $Q_6$ and $Q_8$. In the numerical work we shall use the value of $h$
from Eq.(\ref{cancellation}) and also  $h=5.95$ obtained from the decay width.
%

%With this Lagrangian there are new diagrams for the self energies of the pseudoscalars.
%A quartic divergent term is absorbed into  a universal counter term. The remaining
%momentum independent contributions are combined as mass counter terms. 
We repeated the renormalization procedure and found the following results. For the self energies of 
the pseudoscalars, momentum independent terms combine with the bare masses to define the 
physical masses.  
A momentum dependent
term is included in the wave function renormalization and is the same for $\pi$ and $K$. 
The renormalization of $F_\pi$ and $F_K$ is the same as in ref.\cite{Hambye:PRD}, i.e. there is 
no $h^2$ contribution, which leads to the same value for $L_5$, similarly the value for $L_5-2L_8$
is again very small.  
%%
%We have checked that the quadratic divergences for the factorizable diagrams cancel out. There are
%small corrections remaining for $\langle Q_6\rangle_{0,2}$ and $\langle \pi^0\pi^0|Q_8|K\rangle$. 
%The reason is that they vanish to $\mathcal{O}(p^0)$.  For $\langle \pi^+\pi^-|Q_8|K\rangle$
%the new contribution is also small in comparison with the leading  $\mathcal{O}(p^0)$ term being proportional
%to $r^2 f$.
The quadratic divergences of the factorizable diagrams for  $\langle Q_6\rangle_{0}$,
$\langle Q_6\rangle_{2}$ and $\langle \pi^0\pi^0|Q_8|K\rangle$ cancel out, what remains of 
them are small corrections because to $\mathcal{O}(p^0)$ these matrix elements vanish. The
quadratic divergence from the factorizable diagrams of $\langle \pi^+\pi^-|Q_8|K\rangle$
cancel against the corresponding diagrams with vector meson exchanges when we invoke the condition in
Eq.(\ref{cancellation}). The surviving term is small in comparison with the  $\mathcal{O}(p^0)$
contribution of $\langle \pi^+\pi^-|Q_8|K\rangle$.

We use the following numerical inputs:
\begin{align}
m_\pi &= 0.137 \mbox{GeV},  &m_K &= 0.495\mbox{GeV},   &m_\rho& = 0.771\mbox{GeV}, 
\nonumber\\
f&\equiv F_\pi = 0.0924, &m_s(m_c) &= 0.115\mbox{GeV} ,
&\alpha_S(m_Z) &= 0.117.
\end{align}
The strange quark mass has an error of 0.020 GeV \cite{Buras:03}.
The average quark mass $\hat{m}$ is given by $\hat{m} = m_s/24.4$.
We also use  $\hatlfive = 2.07\times 10^{-3}$, $\hatleight = 1.09\times 10^{-3}$, $Im(\xi_t) = (1.31\pm 0.10)\times 10^{-4}$\cite{Hambye:PRD}
 and the isospin breaking factor of $\Omega_{\eta+\eta'} = 0.15$ \cite{isospinBRK}. 

One can extract $\Lambda^{(4)}_{QCD}$ from $\alpha_S(m_Z)$ at either the
continuum upper limit \cite{PDG}
($\bar{m_b}(\bar{m_b}) = 4.5$ GeV, $\bar{m_c}(\bar{m_c}) = 1.4$ GeV)
or the continuum lower limit
($\bar{m_b}(\bar{m_b}) = 4.0$ GeV, $\bar{m_c}(\bar{m_c}) = 1.0$ GeV):
\begin{align}
\Lambda^{(4)}_{QCD} = 0.279 \pm 0.029 ({\rm upper~limit}),\ \
\Lambda^{(4)}_{QCD} = 0.275 \pm 0.029 ({\rm lower~limit}).
\end{align}
We take, as a conservative estimate, $\Lambda^{(4)}_{QCD} = 0.277\pm 0.031$
GeV ({\em i.e.}, between 0.246 and 0.308 GeV).

%The Wilson coefficients at different regularization scale $\mu$, at
%different levels of precision, and with different regularization schemes,
%can be found in ref \cite{Hambye:NPB}. These values are given for
%$\Lambda^{(4)}_{QCD} = 0.245$ and 0.325 GeV. 
%It is incorrect to interpolate
%%for $\Lambda^{(4)}_{QCD} = 0.308$ GeV since the change is not linear. Even a
%%quadratic interpolation is incorrect. 
%For a first approximation, we just
%use the Wilson coefficients  at 245 and 325 MeV. We use the $\Delta S = 1$  Wilson coefficients  at NLO
%in the  NDR  scheme.

The Wilson coefficients were tabulated \cite{Hambye:NPB} for various
renormalization schemes and the values of $\Lambda^{(4)}_{QCD}$ as functions of the
renormalization scale $\mu$. The values show a convergence among the schemes as
$\mu$ increases and approaches  the value of $\mu=1$ GeV. This is as expected
since QCD is valid at higher momenta.

A second issue is the matching of the coefficients in the various schemes to the
cut-off scale of chiral theory. 
A method for relating the two scales  was suggested in \cite{Bardeen:2001kd}.
The method introduces
%On this topic a regularization in terms of a
%cut-off was suggested \cite{Bardeen:2001kd} by inserting the factor
%
\begin{align}%\label{}
1=\frac{q^2}{q^2-m^2}-\frac{m^2}{q^2-m^2}
\end{align}
and uses the first term as the infrared regulator of QCD and the second term
as the cut-off for the chiral theory. This approach provides a matching of
the two scales $\Lambda_c$ and $\mu$. Recalculation of the evolution of the
coefficients \cite{Bardeen:2001kd} brings the values of the HV scheme closer to
NDR, which are anyway close to the leading order results. All this motivates us
to use the values of the NDR scheme. We shall use values for
$\Lambda^{(4)}_{QCD}=0.245$ GeV, however, we check that interpolation to
$\Lambda^{(4)}_{QCD}=0.277\pm 0.031$ GeV changes the values of 
$\epsilon'/\epsilon$ at most $8\%$. 
%Different  matchings were also introduced in ref.\cite{Wu-match}.
Althernative ways for matching the two theories have also been introduced 
in other articles \cite{Wu-match}.

%\input{sec3}
%%\input{oldSec3}
%\input{sec3new}

%\section{The Cancellation}
{\bf 3. Results.}
As mentioned already,
a previous work demonstrated that renormalization of physical quantities (wave
functions, masses and decay constants) render the factorizable contribution to
$\langle Q_6 \rangle_{0,2}$ and $\langle Q_8 \rangle_{0,2}$ to ${\cal O}(p^0/N_C)$ finite.
%\footnote{We repeated the renormalization scheme of the theory including vector mesons and found that 
%$L^{(r)}_5$ having the same value and $L^{(r)}_8$ having a value close to the one in ref\cite{Hambye:PRD}}.
%
There are loop corrections introduced by the non-factorizable diagrams which to
order $p^0/N_C$ were found to be logarithmic. Going one step further corrections
of order $p^2/N_C$ were studied \cite{Hambye:NPB}, arising from the contact
terms which have a quadratic dependence on the cut-off scale $\Lambda^2_c$.  We
combine the contact terms with the vector meson exchange diagrams and cancel the
quadratic divergence.

%We computed the logarithmic and constant
%terms and present our results in this section.

We present in this section the results for the contact terms and vector meson
exchange diagrams to order $p^2/N_C$ in terms of integrals which are summarized
in Appendix \ref{appendixA}. In order to make the reading easier we give in the
text explicit formulas for the decay $K^0\rightarrow \pi^0\pi^0$ where the
results are shorter. For the decay of $K^0 \rightarrow \pi^+\pi^-$ we collected
the results in Appendix \ref{appendixB}. In both reactions we included the
$\pi^+\pi^-$ and $\pi^0\pi^0$ intermediate states.
%We present the results in terms of
%integrals which are defined in a compact form in the appendix \ref{appendixA}. For reasons of
%brevity we present the results for $K^0\rightarrow \pi^0\pi^0$ where the
%formulae are shorter. For this process the intermediate states can be
%$\pi^+\pi^-$ and $\pi^0\pi^0$. 

The contact terms for $K^0(p_K)\rightarrow \pi^+\pi^-
\rightarrow \pi^0\pi^0$ give
\begin{eqnarray}
  i{\mathcal M}^{00}_{\tmop{con1}} & = & i \frac{2 r^2}{3 \sqrt{2} f^3} [ A
  I_9 ( m_{\pi}, m_{\pi}, p_K, p_K ) + B I_{11} ( m_{\pi}, m_{\pi}, p_K, p_K,
  p_K )\nonumber\\
  &  & - A I_{10} ( m_{\pi}, m_{\pi}, p_K ) - B I_{12} ( m_{\pi}, m_{\pi},
  p_K, p_K )\nonumber\\
  &  & + A C I_8 ( m_{\pi}, m_{\pi}, p_K ) + B C I_9 ( m_{\pi}, m_{\pi},
  p_K, p_K ) ]
\end{eqnarray}
with $A=-8L_5 m_K^2, B=8 L_5, C=(\chi_1+\chi_2)/4+m_K^2-m_\pi^2$ and $\chi_i=r m_i$. 

The contact term for $K^0(p_K)\rightarrow \pi^0\pi^0\rightarrow \pi^0\pi^0$ is 
\begin{eqnarray}
  i{ \mathcal M}^{00}_{\tmop{con2}}& = & i \frac{r^2}{4 \sqrt{2} f^3} C' [ A
  I_8 ( m_{\pi}, m_{\pi}, p_K ) + B I_9 ( m_{\pi}, m_{\pi}, p_K, p_K ) ]
\end{eqnarray}
with $C' = ( \chi_1+ \chi_2 ) $ . The functions $I_i(m_j,m_k,p,\dots)$ etc
represent four dimensional integrals which we define in the
Appendix \ref{appendixA}. The notation with the numbers as subscripts follow the
convention introduced in two Ph.D. theses at Dortmund
University\cite{Kohler+Soldan}, where explicit formulas 
for the functional forms after integration are included.

The $\rho-$exchange diagram for $K^0(p_K)\rightarrow \pi^+\pi^-\rightarrow \pi^0\pi^0$ 
is 
\begin{eqnarray}
  i{\mathcal M}^{00}_{\tmop{exch}1} & = & ( - i ) \frac{2h^2 r^2}{\sqrt{2} f}
  \left\{ - \frac{1}{m_{\rho}^2} [ A I_3 ( m_{\rho}, p_1 ) + B I_4 ( m_{\rho},
  p_1, p_K ) ] \right.\nonumber\\
  &  & + A I_8 ( m_{\pi}, m_{\rho}, p_1 ) + B I_9 ( m_{\pi}, m_{\rho}, p_1,
  p_K )\nonumber\\
  &  & + 2 A I_{30} ( m_{\pi}, m_{\pi}, m_{\rho}, p_K, p_1, p_1 ) + 2 B
  I_{31} ( m_{\pi}, m_{\pi}, m_{\rho}, p_K, p_1, p_K, p_1 )\nonumber\\
  &  & \left. + 2 ( m_K^2 - m_{\pi}^2 ) [ A I_{29} ( m_{\pi}, m_{\pi},
  m_{\rho}, p_K, p_1 ) + B I_{30} ( m_{\pi}, m_{\pi}, m_{\rho}, p_K, p_1, p_K
  ) ] \right\} \nonumber\\ 
\end{eqnarray}
Finally the $\rho-$exchange diagram for $K^0(p_K)\rightarrow \pi^0\pi^0\rightarrow \pi^0\pi^0$ 
is zero
\begin{eqnarray}
  i{\mathcal M}^{00}_{\tmop{exch}2} & = & 0
\end{eqnarray}
because the $\pi^0\pi^0\rho$ vertex does not exist.

%We derived similar results for the decay $K^0\rightarrow \pi^+\pi^-$ and since
%this is a letter we do not include them here but include them in a preprint and
%a longer article \cite{longpaper}.  
Including the vector mesons with the condition in
Eq.(\ref{cancellation}) eliminates the quadratic dependence on the cut-off. This is our method
for regularizing the integrals in terms of physical particles  and interactions which preserve the
symmetries.  The remaining logarithmic dependence of the cut-off will be matched with the
$\ln\mu$ dependence of the QCD.

We give in table \ref{tab1}, the contributions to ${\cal O}(p^2/N_C)$ from the
contact and the $\rho$ exchange terms for $\langle Q_6 \rangle_0$ and $\langle
Q_8 \rangle_2$ in unit of $r^2 \cdot MeV$ as a function of $\Lambda_C$ in the interval $\Lambda_C=0.7$ GeV
to $\Lambda_C=1.0$ GeV. The cut-off scale must be larger than the mass of $\rho$
and the first column is given only as a point of reference. We note that the
dependence of $\langle Q_6\rangle_0^{sum}$ and $\langle Q_8\rangle_2^{sum}$ on
$\Lambda_C$ is very small. Since the value of $h$ from
Eq.(\ref{cancellation}) is smaller than the value obtained from the
$\rho\to\pi\pi$ decay width, we repeated the calculation for $h=5.95$ in table
\ref{tab2}, corresponding to the coupling from $\rho$ decays. The values for
$\epsilon'/\epsilon$ are slightly smaller and the variation of the matrix
elements with the cut-off is larger.  For the calculation of $\rat$ we use, for
the tree and factorizable contributions the values from table I of ref. \cite{Hambye:PRD},
which are primarily responsible for the remaining  $\Lambda_c$ dependence of 
$\rat$.
\begin{table}[htb]
\begin{tabular}{|c|c|c|c|c|}
  \hline
  &   $\Lambda_c = 0.7$ GeV & $\Lambda_c = 0.8$ GeV & $\Lambda_c = 0.9$ GeV & $\Lambda_c = 1.$0 GeV\\
  \hline
  $i \left\langle Q_6 \right\rangle_0^{\tmop{con}}$  & $- 14.8$ & $-17.5$ & $- 20.4$ & $- 23.4$\\
  \hline
  $i \left\langle Q_6 \right\rangle_0^{\rho}$  & 6.5 & 8.9 & 11.6 &14.6\\
  \hline
  $i \left\langle Q_6 \right\rangle_0^{\tmop{sum}}$  & $- 8.3$ & $-8.6$ & $- 8.8$ & $- 8.8$\\
  \hline
& & & & \\
\hline
  $i \left\langle Q_8 \right\rangle_2^{\tmop{con}}$ & 6.24 & 7.43 & 8.7& 10.1\\
  \hline
  $i \left\langle Q_8 \right\rangle_2^{\rho}$ & -2.30 & -3.15 & -4.11& -5.17\\
  \hline
  $i \left\langle Q_8 \right\rangle_2^{\tmop{sum}}$  & 3.94 & 4.28 &4.59 & 4.93\\
  \hline
    &  &  &  & \\
  \hline
  Total $\epsilon' / \epsilon ( 10^{- 3} )$  & 2.23 & 1.84 & 1.53 &1.2\\
  \hline
\end{tabular}
\caption{The contact term and the $\rho-$exchange contributions to $\mathcal{O}(p^2/N_c)$ for the matrix elements of 
 $\langle Q_6 \rangle$ and $\langle Q_8 \rangle$ (in units of $r^2 \cdot MeV$) as well as $\epsilon'/\epsilon$  
as functions of the  cut-off scales $\Lambda_c$.  The value of $h$  is taken from the 
cancellation condition of Eq.(\ref{cancellation}) }\label{tab1}
\end{table}
\begin{table}[htb]
\begin{tabular}{|c|c|c|c|c|}
  \hline
  & $\Lambda_c = 0.7$ GeV & $\Lambda_c = 0.8$ GeV &$\Lambda_c = 0.9$ GeV & $\Lambda_c = 1.$0 GeV\\
  \hline
  $i \left\langle Q_6 \right\rangle_0^{\tmop{con}}$   & $- 14.8$ & $-17.5$ & $- 20.4$ & $- 23.4$\\
  \hline
  $i \left\langle Q_6 \right\rangle_0^{\rho}$   & 9.93 & 13.6 & 17.7 & 22.3\\
  \hline
  $i \left\langle Q_6 \right\rangle_0^{\tmop{sum}}$   & -5.36 &-3.9 & $-2.7$ & -1.1\\
  \hline
  &    &  &  & \\
  \hline
  $i \left\langle Q_8 \right\rangle_2^{\tmop{con}}$   & 6.24 & 7.43 & 8.7& 10.1\\
  \hline
  $i \left\langle Q_8 \right\rangle_2^{\rho}$  & -3.51 & -4.80 & -6.27& -7.89\\
  \hline
  $i \left\langle Q_8 \right\rangle_2^{\tmop{sum}}$  & 2.73 & 2.63 &2.43 & 2.21\\
  \hline
  &    &  &  & \\
  \hline
  Total $\epsilon' / \epsilon ( 10^{- 3} )$  & 2.03 & 1.57 & 1.19 &0.8\\
  \hline
\end{tabular}
\caption{The contact term and the $\rho-$exchange contributions to $\mathcal{O}(p^2/N_c)$ for the  matrix elements of 
 $\langle Q_6 \rangle$ and $\langle Q_8 \rangle$ (in units of $r^2 \cdot MeV$) as well as $\epsilon'/\epsilon$  
as functions of the  cut-off scales $\Lambda_c$.  The value of $h$  is taken  to be the physical one
$h=5.95$.}\label{tab2}
\end{table}
\noindent The results reported in this article present a complete calculation of the matrix elements  $Q_6$ 
and $Q_8$ to order
$p^2/N_C$. The presence of the vector mesons restores to a large extent the unitarity of the theory 
and acts as an upper cut-off for the integrals.
Our results suggest that a non-linear chiral lagrangian with a hidden local symmetry  
may be a more suitable low energy limit for QCD.

%An alternative method has been suggested
%\cite{Bardeen:2001kd} by inserting the factor 
%%
%\begin{align}%\label{}
%1=\frac{q^2}{q^2-m^2}-\frac{m^2}{q^2-m^2}
%\end{align}
%
%and using the first term as the infrared regulator of the QCD term and the
%second providing the cut-off for the chiral theory. The suggestion provides a
%matching of the two scales $\Lambda_C$ and $\mu$. Calculations using the
%infrared term for QCD bring the values of the HV schemes for the coefficient
%functions close to the NDR values. This is a good motivation for using the NDR
%values.

%For $\Lambda^(4)_{QCD}=0.245$ GeV, $m_s(1GeV)=0.125$GeV and Wilson coefficients from
%the NDR scheme we abtain the values for $\epsilon'/\epsilon$ given in the last
%lines of Table. \ref{tab1} and \ref{tab2}. The values are consistent with the experimental
%values in two independent experiments.
As mentioned already, the values of the matrix elements are very stable. 
The calculation
of $\epsilon'/\epsilon$ uses the coefficient functions of NDR at $\Lambda^{(4)}_{QCD}=0.245$GeV and
$m_s(1\mbox{GeV})=0.125$GeV. We found an improved stability of the values for $\epsilon'/\epsilon$ which
are consistent with the experimental results \cite{Exp1,Exp2}.
The main conclusion is that the presence of vector mesons improves the calculation of the matrix elements by making them more
stable functions of the cut-off. 

%The first results are encouraging to continue and complete the calculation. We
%are in the process of calculating the factorizable terms and presenting also the
%constant contributions.  We are studying the effects of the initial state
%interactions and are finally improving the matching with the QCD coefficient by
%improving the calculations in the spirit of Ref.\cite{Hambye:PRD}

We demonstrated that the chiral theory enlarged by the introduction of vector
mesons can eliminate quadratic divergences to $\mathcal{O}{(p^2/N_c)}$. 
The improved stability of $\rat$ is  encouraging to extent the calculation to the
%factorizable terms and 
initial state interactions. We expect the changes to
be small, but we plan to complete them and present them in a longer article \cite{longpaper}.
The extension of the method to the amplitudes $A_0$ and $A_2$ will involve additional operators 
$Q_1, Q_2, \dots$ with considerable increase in the computational work. It will be
interesting, however, to find out whether vector mesons make these amplitudes also 
more stable.    
 
\begin{acknowledgments}
The support of the 
``Bundesministerium f\"ur Bildung, Wissenschaft, Forschung und
Technologie'', Bonn under contract 05HT1PEA9 is gratefully 
acknowledged. A.K. thanks the Alexander von Humboldt-Stiftung for a 
fellowship and the Physics Department of Universit\"at Dortmund (where a 
large part of the work was done) for warm hospitality, and acknowledges
support from the research projects 2000/37/10/BRNS of BRNS, Govt.\ of
India, and F.10-14/2001 (SR-I) of UGC, India.  
Y.F.Z. is grateful to Y.L. Wu for helpful discussions.
\end{acknowledgments}
\appendix
\section{Four dimensional integrals}\label{appendixA}
Several integrals have been used in this article and we try to define then in a compact notation. 
The integrals $I_3$, $I_4$ have the same denominator but have different numerators separated from
each other with semicolons
\begin{align}%\label{}
I_{3;4}=\frac{i}{(2\pi)^4}\int d^4 q \frac{\{1;(p\cdot q)\}}{(q-k)^2-m^2}
\end{align}
The integrals $I_8$, $I_9$, $I_{10}$, $I_{11}$ and $I_{12}$ have again the same
denominator but have different numerators separated from each other with
semicolons
\begin{align}%\label{}
I_{8;9;10;11;12}=\frac{i}{(2\pi)^4}\int d^4 q \frac{\{1;(p\cdot q); q^2;(p_1\cdot q)(p_2\cdot q);q^2(p\cdot q)\}}
{(q^2-m_1^2)[(q-k)^2-m_2^2]}
\end{align}
The same notation is used in the integrals $I_{29}$, $I_{30}$ and $I_{31}$,  
\begin{align}%\label{}
I_{29;30;31}=\frac{i}{(2\pi)^4}\int d^4 q \frac{\{1;(p_1\cdot q); (p_1\cdot q)(p_2\cdot q)\}}
{(q^2-m_1^2)[(q-k)^2-m_2^2][(q-p)^2-m_3^2]}
\end{align}
Among these integrals $I_3$, $I_4$, $I_{10}$,$I_{11}$ and $I_{12}$ have  quadratic divergences in the 
cut-off regularization scheme. The quadraticlly divergent parts are given by
\begin{align}%\label{}
I_3(m,k)|_{\Lambda_c^2div} &=\frac{1}{(4\pi)^2} \Lambda_c^2 
\nonumber \\
I_4(m,k,p)|_{\Lambda_c^2div} &=\frac{(k\cdot p)}{2(4\pi)^2} \Lambda_c^2 
\nonumber \\
I_{10}(m_1,m_2,k)|_{\Lambda_c^2div} &=\frac{1}{(4\pi)^2} \Lambda_c^2 
\nonumber \\
I_{11}(m_1,m_2,k,p_1,p_2)|_{\Lambda_c^2div} &=\frac{(p_1\cdot p_2)}{4(4\pi)^2} \Lambda_c^2 
\nonumber \\
I_{12}(m_1,m_2,k,p)|_{\Lambda_c^2div} &=\frac{(k\cdot p)}{2(4\pi)^2} \Lambda_c^2 
\end{align}
Using the quadratic divergences and the formulas in this article the reader can verify the cancellations. 
\section{$K\to \pi^+\pi^-$ decay amplitudes at $\mathcal{O}(p^2/N_c)$}\label{appendixB}
The contact term for  $K^0 ( p_K ) \rightarrow \pi^+ \pi^- \rightarrow\pi^+ ( p_1 )
\pi^- ( p_2 )$ is given by
\begin{align}
  i\mathcal{M}_{\tmop{con} 1}^{+ -} & =  - i \frac{r^2}{3 \sqrt{2} f^3} [ A
  I_{10} ( m_{\pi}, m_{\pi}, p_K ) + B I_{12} ( m_{\pi}, m_{\pi}, p_K, p_K )
\nonumber\\
  &   - 2 A I_9 ( m_{\pi}, m_{\pi}, p_K, 2 p_2 - p_1 ) - 2 B_{} I_{11} (
  m_{\pi}, m_{\pi}, p_K, p_{K,} 2 p_2 - p_1 )
\nonumber\\
  &   - A C I_8 ( m_{\pi}, m_{\pi}, p_K ) - B C I_9 ( m_{\pi}, m_{\pi}, p_K,
  p_K ) ]
\end{align}
with $C_{} = ( \chi_1 + \chi_2 ) + ( m_K^2 - m_{\pi}^2 )$.

The contact term for $K^0 ( p_K ) \rightarrow \pi^0 \pi^0 \rightarrow
\pi^+ ( p_1 ) \pi^- ( p_2 )$ is 
\begin{align}
  i\mathcal{M}_{\tmop{con} 2}^{+ -} & =  - i \frac{r^2}{3 \sqrt{2} f^3} [ A
  I_{10} ( m_{\pi}, m_{\pi}, p_K ) + B I_{12} ( m_{\pi}, m_{\pi}, p_K, p_K )
\nonumber\\
  &   - A I_9 ( m_{\pi}, m_{\pi}, p_K, p_K ) - B_{} I_{11} ( m_{\pi},
  m_{\pi}, p_K, p_{K,} p_K )
\nonumber\\
  &   - A C' I_8 ( m_{\pi}, m_{\pi}, p_K ) - B C' I_9 ( m_{\pi}, m_{\pi},
  p_K, p_K ) ]
\end{align}

The $\rho -$exchange diagram  through $\pi^+ \pi^- \tmop{loop}$ gives 
\begin{align}
  i\mathcal{M}_{\tmop{exch} 1}^{+ -} & = ( - i ) \frac{h^2 r^2}{\sqrt{2} f}
  \left\{ - \frac{1}{m_{\rho}^2} [ A I_3 ( m_{\rho}, p_1 ) + B I_4 ( m_{\rho},
  p_1, p_K ) ] \right.
\nonumber\\
  &  + A I_8 ( m_{\pi}, m_{\rho}, p_1 ) + B I_9 ( m_{\pi}, m_{\rho}, p_1,
  p_K )
\nonumber\\
  &   + 2 A I_{30} ( m_{\pi}, m_{\pi}, m_{\rho}, p_K, p_1, p_1 ) + 2 B
  I_{31} ( m_{\pi}, m_{\pi}, m_{\rho}, p_K, p_1, p_K, p_1 )
\nonumber\\
  &   \left. + 2 ( m_K^2 - m_{\pi}^2 ) [ A I_{29} ( m_{\pi}, m_{\pi},
  m_{\rho}, p_K, p_1 ) + B I_{30} ( m_{\pi}, m_{\pi}, m_{\rho}, p_K, p_1, p_K
  ) ] \right\}
\end{align}
with $C' = ( \chi_1 + \chi_2 ) / 4 + ( m_K^2 - m_{\pi}^2 )$.

The $\rho -$exchange diagram through $\pi^0 \pi^0 \tmop{loop}$ gives the 
same contribution, i.e. 
\begin{align}
\mathcal{M}_{\tmop{exch} 2}^{+ -} & =  \mathcal{M}_{\tmop{exch} 1}^{+ -} 
\end{align}

It is straight forward to verify that the cancellation condition of Eq.(\ref{cancellation})
also holds for $K\to \pi^+\pi^-$.  
%\newpage


\begin{thebibliography}{54}
%\setlength{\parskip}{0.15cm}
%\setlength{\baselineskip}{0.35cm}
\bibitem{Hambye:PRD} T. Hambye,G.O.Kohler,E.A.Paschos, P.H.Soldan and W.A. Bardeen, \PRD(58,014017,1998).  %Hambye:PRD
\bibitem{ref1} A.~Buras et al.,
        {\it Nucl.\ Phys.} {\bf B400} (1993) 37
\bibitem{ref2} M.~Ciuchini et al.,
        {\it Nucl.\ Phys.} {\bf B415} (1994) 403
\bibitem{BBG}W.A. Bardeen. A.J. Buras, and J.-M.Gerard, Nucl.Phys.{\bf B293},787(1987)
Phys. Lett {\bf B192},138(1987),{\bf B211},343,1988
\bibitem{Hambye:NPB} T.~Hambye, G.O. Kohler, E.A. Paschos, P.H. Soldan %Hambye:NPB
        {\it Nucl.\ Phys.} {\bf B564} (2000) 391 
\bibitem{Bando}
M.~Bando, T.~Kugo, S.~Uehara, K.~Yamawaki and T.~Yanagida,
%``Is Rho Meson A Dynamical Gauge Boson Of Hidden Local Symmetry?,''
Phys.\ Rev.\ Lett.\  {\bf 54}, 1215 (1985).
%%CITATION = PRLTA,54,1215;%%
M.~Bando, T.~Kugo and K.~Yamawaki,
%``Nonlinear Realization And Hidden Local Symmetries,''
Phys.\ Rept.\  {\bf 164}, 217 (1988).
%%CITATION = PRPLC,164,217;%%
\bibitem{ref5} O.~Kaymakcalan and J.~Schechter,
        {\it Phys.\ Rev.} {\bf D31} (1985) 1109 .
U.~G.~Meissner,
%``Low-Energy Hadron Physics From Effective Chiral Lagrangians With Vector
%Mesons,''
Phys.\ Rept.\  {\bf 161}, 213 (1988).
%%CITATION = PRPLC,161,213;%%
\bibitem{ref5b}
Fayyazuddin and Riazuddin,
%``Effective Chiral Lagrangian Of Pseudoscalar And Vector Mesons,''
Phys.\ Rev.\ D {\bf 36}, 2768 (1987).
%%CITATION = PHRVA,D36,2768;%%
G.~Isidori and A.~Pugliese,
%``Chiral weak Lagrangian for vector mesons and K $\to$ 3 pi decay amplitudes,''
Nucl.\ Phys.\ B {\bf 385}, 437 (1992).
%%CITATION = NUPHA,B385,437;%%
J.~Schechter and A.~Subbaraman,
%``Role of light vector mesons in the heavy particle chiral Lagrangian,''
Phys.\ Rev.\ D {\bf 48}, 332 (1993)
[arXiv:hep-ph/9209256].
%%CITATION = HEP-PH 9209256;%%
\bibitem{longpaper} E.A.~Paschos et al.  in preparation.
\bibitem{gasser-leutwyler}J.~Gasser and H. Leutwyler. Nucl. Phys. {\bf B250},465(1985)\
{\bf B250}, 517(1985){\bf B250}, 539(1985)
        {\it } {\bf } 
%\cite{Bardeen:2001kd}
\bibitem{epsilonP}
G.~Buchalla, A.~J.~Buras and M.~K.~Harlander,
%``The Anatomy Of Epsilon-Prime / Epsilon In The Standard Model,''
Nucl.\ Phys.\ B {\bf 337}, 313 (1990).
%%CITATION = NUPHA,B337,313;%%
A.~J.~Buras, M.~Jamin and M.~E.~Lautenbacher,
%``The Anatomy of epsilon-prime / epsilon beyond leading logarithms with
%improved hadronic matrix elements,''
Nucl.\ Phys.\ B {\bf 408}, 209 (1993).
%[arXiv:hep-ph/9303284].
%%CITATION = HEP-PH 9303284;%%
S.~Bosch, A.~J.~Buras, M.~Gorbahn, S.~Jager, M.~Jamin, M.~E.~Lautenbacher and L.~Silvestrini,
%``Standard model confronting new results for epsilon'/epsilon,''
Nucl.\ Phys.\ B {\bf 565}, 3 (2000).
%[arXiv:hep-ph/9904408].
%%CITATION = HEP-PH 9904408;%
S.~Bertolini, J.~O.~Eeg and M.~Fabbrichesi,
%``An updated analysis of epsilon'/epsilon in the standard model with  hadronic
%matrix elements from the chiral quark model,''
Phys.\ Rev.\ D {\bf 63}, 056009 (2001).
%[arXiv:hep-ph/0002234].
%%CITATION = HEP-PH 0002234;%%
E.~Pallante, A.~Pich and I.~Scimemi,
%``The standard model prediction for epsilon'/epsilon,''
Nucl.\ Phys.\ B {\bf 617}, 441 (2001).
%[arXiv:hep-ph/0105011].
%%CITATION = HEP-PH 0105011;%%
E.~A.~Paschos and Y.~L.~Wu,
%``Correlations Between Epsilon-Prime / Epsilon And Heavy Top,''
Mod.\ Phys.\ Lett.\ A {\bf 6}, 93 (1991).
%%CITATION = MPLAE,A6,93;%%
J.~Heinrich, E.~A.~Paschos, J.~M.~Schwarz and Y.~L.~Wu,
%``Accuracy Of The Predictions For Direct CP Violation,''
Phys.\ Lett.\ B {\bf 279}, 140 (1992).
%%CITATION = PHLTA,B279,140;%%
A.~A.~Belkov, G.~Bohm, A.~V.~Lanyov and A.~A.~Moshkin,
%``Phenomenological analysis of epsilon'/epsilon within an effective  chiral
%Lagrangian approach at O(p**6),''
arXiv:hep-ph/9907335.
%%CITATION = HEP-PH 9907335;%%
See also ref. \cite{Hambye:NPB,Wu-match}.
%
%
\bibitem{KSRF}
K.~Kawarabayashi and M.~Suzuki,
%``Partially Conserved Axial Vector Current And The Decays Of Vector Mesons,''
Phys.\ Rev.\ Lett.\  {\bf 16}, 255 (1966).
%%CITATION = PRLTA,16,255;%%
Riazuddin and Fayyazuddin,
%``Algebra Of Current Components And Decay Widths Of Rho And K* Mesons,''
Phys.\ Rev.\  {\bf 147}, 1071 (1966).
%%CITATION = PHRVA,147,1071;%%
\bibitem{Buras:03}
A.~J.~Buras and M.~Jamin,
%``Epsilon'/epsilon at the NLO: 10 years later,''
JHEP {\bf 0401}, 048 (2004)
%[arXiv:hep-ph/0306217].
%%CITATION = HEP-PH 0306217;%%
\bibitem{isospinBRK}
There are several articles on the isospin-breaking term which give a wide range of 
values. e.g.
A.~J.~Buras and J.~M.~Gerard,
%``Epsilon-Prime / Epsilon In The Standard Model,''
Phys.\ Lett.\ B {\bf 192}, 156 (1987).
%%CITATION = PHLTA,B203,272;%%
S.~Gardner and G.~Valencia,
%``Additional isospin-breaking effects in $\epsilon^\prime/\epsilon$,''
Phys.\ Lett.\ B {\bf 466}, 355 (1999).
%[arXiv:hep-ph/9909202].
%%CITATION = HEP-PH 9909202;%%
G.~Ecker, G.~Muller, H.~Neufeld and A.~Pich,
%``pi0 eta mixing and CP violation,''
Phys.\ Lett.\ B {\bf 477}, 88 (2000).
%[arXiv:hep-ph/9912264].
%%CITATION = HEP-PH 9912264;%%
C.~E.~Wolfe and K.~Maltman,
%``Strong isospin-breaking effects on the Delta(I) = 3/2 amplitude in  K $\to$
%2pi at next-to-leading order in the chiral expansion,''
Phys.\ Lett.\ B {\bf 482}, 77 (2000).
%[arXiv:hep-ph/9912254].
%%CITATION = HEP-PH 9912254;%%
S.~Gardner and G.~Valencia,
%``The impact of $|$Delta(I)$|$ = 5/2 transitions in K $\to$ pi pi decays,''
Phys.\ Rev.\ D {\bf 62}, 094024 (2000).
%[arXiv:hep-ph/0006240].
%%CITATION = HEP-PH 0006240;%%
V.~Cirigliano, A.~Pich, G.~Ecker and H.~Neufeld,
%``Isospin violation in epsilon',''
Phys.\ Rev.\ Lett.\  {\bf 91}, 162001 (2003).
%[arXiv:hep-ph/0307030].
%%CITATION = HEP-PH 0307030;%%
V.~Cirigliano, G.~Ecker, H.~Neufeld and A.~Pich,
%``Isospin breaking in K $\to$ pi pi decays,''
Eur.\ Phys.\ J.\ C {\bf 33}, 369 (2004).
%[arXiv:hep-ph/0310351].
%%CITATION = HEP-PH 0310351;%%
\bibitem{PDG} Particle Data Group, K.Hagiwara et al., Phys. Rev. {\bf D}66,010001(2002).
\bibitem{Bardeen:2001kd}
W.~A.~Bardeen,
%``On the large N(c) expansion in quantum chromodynamics,''
Nucl. Phys. {\bf B} ,suppl. proc. Supp. 7A, 149,(1989), 
Fortsch.\ Phys.\  {\bf 50} (2002) 483
[arXiv:hep-ph/0112229].
%%CITATION = HEP-PH 0112229;%%
\bibitem{Wu-match}
Y.~L.~Wu,
%``A new prediction for direct CP violation epsilon'/epsilon and Delta(I)  = 1/2
%rule,''
Phys.\ Rev.\ D {\bf 64}, 016001 (2001)
%\bibitem{Peris:2003gw}
T.~Hambye, S.~Peris and E.~de Rafael,
%``Delta(I) = 1/2 and epsilon'/epsilon in large-N(c) QCD,''
JHEP {\bf 0305}, 027 (2003)
[arXiv:hep-ph/0305104].
%%CITATION = HEP-PH 0305104;%%
S.~Peris,
%``Unfactorizing polychromatic penguins,''
arXiv:hep-ph/0310063.
%%CITATION = HEP-PH 0310063;%%
%\bibitem{Bijnens:2000im}
J.~Bijnens and J.~Prades,
%``epsilon'(K)/epsilon(K) in the chiral limit,''
JHEP {\bf 0006}, 035 (2000)
[arXiv:hep-ph/0005189].
%%CITATION = HEP-PH 0005189;%%
%\bibitem{Bijnens:2003hk}
J.~Bijnens, E.~Gamiz and J.~Prades,
%``Hadronic matrix elements for kaons,''
arXiv:hep-ph/0309216.
%%CITATION = HEP-PH 0309216;%%

%[arXiv:hep-ph/0012371].
%%CITATION = HEP-PH 0012371;%%
\bibitem{Kohler+Soldan}G.O. Kohler and P.H. Soldan, Ph.D. theses (1998). Dortmund University.
\bibitem{Exp1} V.Fanti et al., Phys.Lett.{B 465},335 (1999).
\bibitem{Exp2} A.Alavi-Harati et al., Phys. Rev. Lett,83,22 (1999).
\end{thebibliography}
\end{document}